\documentclass{article}

\usepackage{spconf,amsmath,amssymb,graphicx}
\usepackage{booktabs}
\usepackage{cite}
\usepackage{multirow}
\usepackage{placeins}
\usepackage{url}
\usepackage{enumitem}
\usepackage[table]{xcolor}
\usepackage[hidelinks]{hyperref}
\usepackage{orcidlink}

\definecolor{oursbg}{HTML}{EAF2F8}
\definecolor{crimson}{HTML}{8B0000}
\newcommand{\best}[1]{\textcolor{crimson}{\textbf{#1}}}
\newcommand{\second}[1]{\underline{#1}}

\makeatletter
\long\def\@makecaption#1#2{%
  \vskip 3pt
  \setbox\@tempboxa\hbox{#1. #2}%
  \ifdim \wd\@tempboxa >\hsize
    #1. #2\par
  \else
    \hbox to\hsize{\hfil\box\@tempboxa\hfil}%
  \fi}
\makeatother

\graphicspath{{figures/}}

\title{RadioTraj3D: Height-Aware Adaptive Sampling for 3D Radio-Map Reconstruction from Ultra-Sparse Trajectories}

\name{
    \normalsize
    Zixuan Guo\,$^{1}$\orcidlink{0009-0000-1171-252X},
    Xiucheng Wang\,$^{2}$\orcidlink{0000-0003-1439-4875},
    Wenshuo Chen\,$^{3}$\orcidlink{0009-0002-1966-6059},
    Conghao Zhou\,$^{2}$\orcidlink{0000-0002-5727-2432},
    Chuan Huang\,$^{4}$\orcidlink{0000-0001-5965-0823},
    Ruijin Sun\,$^{2}$\orcidlink{0000-0002-4403-9893},
    Nan Cheng\,$^{2}$\sthanks{
        Corresponding author is Nan Cheng
        (e-mail: dr.nan.cheng@ieee.org).
    }\orcidlink{0000-0001-7907-2071},
}

\address{
    \small
    \shortstack{
        $^{1}$School of Physics,
        Xidian University, Xi'an 710071, China\\[2pt]
        $^{2}$School of Telecommunications Engineering,
        Xidian University, Xi'an 710071, China\\[2pt]
        $^{3}$Artificial Intelligence Thrust, Information Hub,
        HKUST (GZ), Guangzhou 511453, China\\[2pt]
        $^{4}$Shenzhen Institute for Advanced Study,
        University of Electronic Science and Technology of China,
        Shenzhen 518110, China
    }
}

\begin{document}
\ninept
\setlength{\parskip}{0pt}
\setlength{\parindent}{1em}
\setlength{\textfloatsep}{3pt plus 1pt minus 1pt}
\setlength{\floatsep}{3pt plus 1pt minus 1pt}
\setlength{\intextsep}{3pt plus 1pt minus 1pt}
\setlist[enumerate]{leftmargin=1.7em,label=\arabic*.,topsep=2pt,
  itemsep=1pt,parsep=0pt,partopsep=0pt}

\maketitle

\begin{abstract}
Aerial networks require radio propagation information at unvisited positions and heights to plan routes, select links, and allocate wireless resources. A three-dimensional radio map (3D RM) represents propagation conditions over the service area, enabling these decisions before flight. However, constructing such a map from autonomous aerial vehicle (AAV) measurements is challenging, since radio measurements are available only along the trajectory, whereas the target covers the 3D service volume. Existing methods use fixed AAV trajectories that ignore how building distributions and heights affect electromagnetic propagation, missing the heights most informative for RM reconstruction. To address these issues, we propose RadioTraj3D, which uses building-height information to coordinate measurement selection and layer-wise reconstruction. Building-height variations reveal propagation-sensitive layers that guide height-adaptive sampling. We therefore propose G-METS to select geometry-sensitive heights and allocate an exact budget across edge, corridor, and open regions before channel acquisition. A dual-view reconstructor then integrates local three-dimensional boundaries with full-height context through one bounded coefficient per layer. Simulated outdoor and indoor maps demonstrate state-of-the-art reconstruction across environments and vertical grids. On real-world AAV measurements from the LACN dataset, RadioTraj3D improves peak signal-to-noise ratio by 0.84 dB and reduces normalized mean squared error by 17.5\% over Completion AE.
\end{abstract}


\section{Introduction}
\label{sec:introduction}

Aerial route planning, link selection, and wireless resource allocation require radio propagation information at positions and heights not visited by an autonomous aerial vehicle (AAV). A three-dimensional radio map (3D RM) represents radio propagation conditions across the service volume for preflight decisions \cite{wang2026tutoriallearningbasedradiomap}. When transmitter coordinates and antenna settings are unavailable in shared-spectrum deployments, the map must be reconstructed from radio measurements. Map errors can direct an AAV toward weak connectivity, underestimate interference, or reject viable routes. Exhaustive radio measurements consume flight time, while measurements at a few heights leave most layers unobserved \cite{shrestha2023spectrum,hu2023constrained}. We therefore study transmitter-agnostic 3D RM reconstruction from sparse radio measurements collected along feasible AAV trajectories \cite{hu2023constrained,qiu2026trajectory}.

Full 3D reconstruction from trajectory-constrained radio measurements faces the dual challenges of selecting informative locations before flight and recovering unobserved heights after collection. Before collection, locations must form a collision-free path and meet an exact budget without known propagation patterns. Fixed trajectories and heights ignore building-height variations, so they can miss propagation-sensitive layers and major blockage changes \cite{hu2023constrained,zhao2025threedradiodiff,wang2026radiodiff3d}. Random masks also create layouts infeasible for continuous AAV paths \cite{qiu2026trajectory}. The resulting sparse measurements must capture local blockage and vertical variation across the unobserved volume. Their locations therefore govern spatial coverage and vertical observability. Planar estimators reconstruct fixed-height maps \cite{levie2021radiounet,teganya2022deepcompletion}, 3D methods require transmitter information \cite{zhao2025threedradiodiff,wang2026radiodiff3d}, and trajectory-aware methods do not recover the complete 3D RM \cite{shrestha2023spectrum,hu2023constrained,qiu2026trajectory}. Consequently, existing methods cannot jointly select informative locations before flight and reconstruct unobserved heights without transmitter information.

\begin{figure}[!t]
\centering
\includegraphics[width=\linewidth]{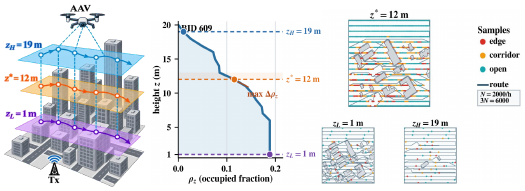}
\caption{G-METS collision-free exact-budget sampling.}
\label{fig:sampling_framework}
\end{figure}

Building-height variations change blockage patterns across adjacent layers and are available before radio measurements are collected. They reveal propagation-sensitive layers for height-adaptive sampling. The same height dependence changes the relative usefulness of local 3D boundaries and full-height context during reconstruction. To exploit both relations, we propose RadioTraj3D, which coordinates measurement selection and layer-wise reconstruction. G-METS selects geometry-sensitive heights and allocates the exact measurement budget across edge, corridor, and open regions. The dual-view reconstructor integrates local 3D boundaries with full-height context through one bounded coefficient per layer. The main contributions are summarized below.
\begin{enumerate}
    \item We formulate transmitter-agnostic 3D RM reconstruction from sparse radio measurements collected along collision-free AAV trajectories under an exact measurement budget, separating building information available before flight from measurements collected during flight.
    \item We propose G-METS to select geometry-sensitive heights from building-height variations and allocate the measurement budget across edge, corridor, and open regions while limiting nearby redundancy.
    \item We design a dual-view reconstructor that combines local 3D boundaries with full-height context and adapts their contributions through one bounded coefficient per layer.
    \item We demonstrate state-of-the-art reconstruction across simulated outdoor and indoor radio maps with different vertical grids. On real-world radio measurements collected by AAVs in the low-altitude communication network (LACN) dataset~\cite{huang2026lacn}, the RadioTraj3D reconstructor outperforms Completion AE in peak signal-to-noise ratio at every withheld height.
\end{enumerate}

\section{System Model}
\label{sec:system_model}

To formalize the coupled task, let $\mathcal{Y}\in[0,1]^{D\times H\times W}$ denote the normalized 3D pathloss RM and let $\mathcal{B}\in\{0,1\}^{D\times H\times W}$ denote its occupancy grid, where one marks an obstruction. The layout is known, but transmitter coordinates, antenna patterns, and other source data are unavailable. We assume calibrated pathloss measurements with a known transmit level. The formulation also applies to received-power maps when this level is unknown.

Within this input model, let $\mathcal{I}=\{1,\ldots,D\}$ index receiver heights and let $\mathcal{I}_{\mathrm{o}}\subset\mathcal{I}$ denote the observed heights. At $z_i$, $\mathcal{P}_i$ is an ordered collision-free subroute, and $\mathcal{S}_i\subseteq\mathcal{P}_i$ contains exactly $N$ unique samples. The observation mask $\mathcal{M}$ and sparse tensor $\mathcal{V}$ are
\begin{align}
  \mathcal{M}_{i,x,y}
  &= \begin{cases}
       1, & i\in\mathcal{I}_{\mathrm{o}},\ (x,y)\in\mathcal{S}_i,\\
       0, & \text{otherwise},
     \end{cases}
  \label{eq:observation_mask}\\
  \sum_{x,y}\mathcal{M}_{i,x,y} &= N,
  \quad \forall\,i\in\mathcal{I}_{\mathrm{o}},
  \label{eq:exact_budget}\\
  \mathcal{V} &= \mathcal{M}\odot\mathcal{Y}.
  \label{eq:sparse_observation}
\end{align}
These constraints define the pre-flight information contract. G-METS determines $\mathcal{I}_{\mathrm{o}}$, $\mathcal{P}_i$, and $\mathcal{S}_i$ only from $\mathcal{B}$ and the route constraints. It connects planar subroutes above nearby obstacles and includes vertical transfers in the 3D route length. The complete route and sample budget are therefore fixed before any pathloss value is read.

After acquisition, the reconstructor receives $\mathcal{B}$, $\mathcal{V}$, $\mathcal{M}$, a height encoding $\mathcal{Z}$, and a deterministic coarse estimate $\mathcal{C}(\mathcal{V},\mathcal{M})$. This estimate supplies mask-normalized context at several scales and initializes unsupported voxels with the observed mean. For parameters $\theta$, the output is
\begin{align}
  \widehat{\mathcal{Y}}
  &= f_{\theta}(\mathcal{B},\mathcal{V},\mathcal{M},\mathcal{Z},\mathcal{C}).
  \label{eq:reconstruction_model}
\end{align}
The complete 3D RM is the learning target, and free-space mean squared error (MSE) and boundary errors guide training. This formulation makes the coupling explicit. The route determines which 3D structures become observable, and the reconstructor must turn that evidence into the remaining volume.

\section{Proposed Method}
\label{sec:method}

RadioTraj3D follows this information order. Before reading field values, G-METS chooses the observation heights and places exactly $N$ samples on each obstacle-avoiding subroute. The dual-view network then reconstructs the complete RM from geometry and sparse measurements without transmitter data. This sequence preserves route feasibility and aligns reconstruction with evidence that the route can actually collect.

\subsection{G-METS Sampling}

G-METS makes the pre-flight sample set informative by adapting observation heights to the building layout. If $\rho_i$ is the occupied-area fraction at $z_i$, the strongest roof transition is
\begin{align}
  i^{\star}=\operatorname*{arg\,max}_{i\in\{1,\ldots,D-1\}}
  \left(\rho_i-\rho_{i+1}\right).
  \label{eq:roof_transition}
\end{align}

\begin{figure}[!t]
\centering
\includegraphics[width=\linewidth]{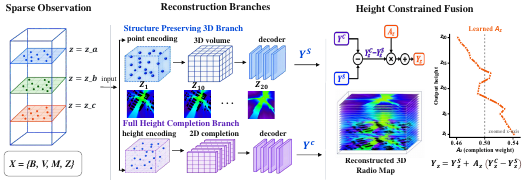}
\caption{RadioTraj3D dual-view reconstruction.}
\label{fig:reconstruction_framework}
\end{figure}
The selected middle height complements fixed low and high anchors, covering near-ground, roof-transition, and above-roof propagation. At each height, G-METS builds a collision-free serpentine path and groups candidates into edge, corridor, and open regions. These regions target shadow changes, obstructed streets, and broad spatial coverage. A greedy allocator favors uncovered cells and avoids nearby repeats subject to regional availability. If the preferred spacing cannot fill the budget, G-METS reduces it until exactly $N$ points are selected. Height selection exposes vertical transitions, while regional allocation spreads the planar budget across propagation regimes. Together, they provide evidence near roof changes and horizontal obstruction boundaries. Because the height mapping and thresholds are fixed from training geometry, G-METS adapts to each layout without reading test pathloss values.

\subsection{Complementary Reconstruction Branches}

Once measurements are collected, the two branches address complementary effects of missing data, as shown in Fig.~\ref{fig:reconstruction_framework}. The structure-preserving branch injects point and building tokens into a shared-weight 3D U-Net to retain local boundaries. The completion branch stacks height as channels in a mask-aware 2D encoder and decoder to supply full-height context. Both receive the same inputs and coarse estimate, so their difference lies in how they organize spatial and vertical information.
\begin{align}
  \mathcal{Y}^{S} &= f_S(\mathcal{B},\mathcal{V},\mathcal{M},\mathcal{Z},\mathcal{C}),
  \label{eq:native_predictor}\\
  \mathcal{Y}^{C} &= f_C(\mathcal{B},\mathcal{V},\mathcal{M},\mathcal{Z},\mathcal{C}).
  \label{eq:completion_predictor}
\end{align}
The relative accuracy of the two branches changes with altitude because boundary visibility and measurement support also vary with height. RadioTraj3D therefore assigns one bounded coefficient to each output layer. For $i\in\mathcal{I}$,
\begin{align}
  A_i &= A_{\max}\,\operatorname{sigmoid}(a_i), \qquad 0<A_i<A_{\max},
  \label{eq:height_coefficient}\\
  \widehat{\mathcal{Y}}_i
  &= (1-A_i)\mathcal{Y}^{S}_{i}+A_i\mathcal{Y}^{C}_{i},
  \label{eq:bounded_fusion}
\end{align}
Each $A_i$ is shared across layouts and positions, so it represents a height-specific preference without learning a gate for every voxel. This spatial sharing matches sparse trajectories, which do not provide enough local evidence for a separate gate at every position. Equation~\eqref{eq:bounded_fusion} also keeps each estimate between the two branch predictions. With both branches frozen, the coefficients minimize an objective whose free-space term is mean squared error (MSE). The complete objective is
\begin{align}
\mathcal{L}={}&\mathcal{L}_{\rm free}^{\rm MSE}
+\lambda_{\rm bnd}\mathcal{L}_{\rm bnd}^{\rm L1}
+\lambda_{\rm grad}\mathcal{L}_{\rm grad}^{\rm L1}
+\lambda_A\|A-A_0\|_2^2,
\end{align}
where the terms measure free-space, boundary, and gradient errors together with distance from the initial weights. Calibration requires only $D$ parameters instead of $DHW$ voxel gates, allowing the branch balance to change with height at low cost. The bound preserves each output as an interpolation between the two views, so the coefficient profile records their relative reliability across altitude.

\FloatBarrier
\section{Experiments}
\label{sec:experiments}

\subsection{Experimental Setup}

The evaluation separates reconstruction, fusion, and acquisition through matched observations, controlled ablations, and frozen-model sampler comparisons. Transfer and density tests vary the environment, vertical grid, measurement support, and data source. UrbanRadio3D~\cite{wang2026radiodiff3d} contains normalized $20\!\times\!256\!\times\!256$ pathloss tensors generated at 5.9~GHz by the WinProp/FEKO Dominant Path Model. Each urban scenario is indexed by a building identifier (BID). We train on 600 scenarios and test on another 50 non-overlapping scenarios. Paired statistics use all the 50 test scenarios because complete method pairs are available for those cases. All maps are normalized from $[-169,-92]$~dB to $[0,1]$. G-METS anchors 1 and 19~m, maps $i^\star$ to 12, 13, or 14~m from training geometry, and allocates 25\%, 25\%, and 50\% of samples to edge, corridor, and open regions. Tests use $N\in\{500,1000,2000\}$. Safe uses equal arc-length sampling, while geometry-aware coverage adds regional quotas and farthest-point selection. All samplers share the same route and budget constraints. The three-stage shared-weight U-Net maps 128-D point and building tokens to 32 features and uses fixed multiscale box filters for coarse context. We set $A_{\max}=0.8$, $A_0=0.5$, $\lambda_{\rm bnd}=\lambda_{\rm grad}=0.05$, and $\lambda_A=10^{-3}$. Both branches train for 400 updates and then freeze while $A_i$ is calibrated for 300 updates, isolating the fusion rule. Following RadioDiff and RadioDiff-3D~\cite{wang2025radiodiff,wang2026radiodiff3d}, we report PSNR, NMSE, structural similarity index measure (SSIM), and root mean squared error (RMSE). PSNR, NMSE, and RMSE cover the normalized volume, while SSIM is averaged across height slices. All reconstruction methods use the same test cases, transmitter-agnostic inputs, pre-measurement G-METS coordinates, target volume, evaluator, and training budget. For sampler tests, three equally initialized completion models train on Safe, geometry-aware, or G-METS samples and then freeze for cross-evaluation. Component and height ablations reuse one frozen G-METS model. These controls isolate reconstruction and sampling gains without extra field information.

\FloatBarrier
\subsection{Protocol-Matched Reconstruction Comparison}

\begin{table}[!t]
\centering
\caption{Matched reconstruction on UrbanRadio3D.}
\label{tab:external_baselines}
\ninept
\setlength{\tabcolsep}{2.6pt}
\renewcommand{\arraystretch}{0.9}
\begin{tabular}{@{}lcccc@{}}
\toprule
Method & PSNR $\uparrow$ & SSIM $\uparrow$ & NMSE $\downarrow$ & RMSE $\downarrow$ \\
\midrule
IDW~\cite{shepard1968interpolation} & 19.18 & 0.685 & 0.0608 & 0.1115 \\
Sparse 3D U-Net~\cite{wang2026radiodiff3d} & 20.61 & 0.717 & 0.0451 & 0.1053 \\
Completion AE~\cite{teganya2022deepcompletion} & \second{21.28} & \second{0.744} & \second{0.0396} & \second{0.0991} \\
RadioDiff-3D~\cite{wang2026radiodiff3d} & 11.06 & 0.214 & 0.4043 & 0.2824 \\
RadioLAM~\cite{liu2026radiolam} & 7.18 & 0.460 & 0.9898 & 0.4413 \\
RadioDUN~\cite{chen2026radiodun} & 11.70 & 0.258 & 0.3017 & 0.2622 \\
DAT-UNet~\cite{liu2026datunet} & 17.30 & 0.556 & 0.0878 & 0.1373 \\
\rowcolor{oursbg}\textbf{RadioTraj3D} & \best{23.86} & \best{0.805} & \best{0.0247} & \best{0.0664} \\
\midrule
\rowcolor{gray!12}\textbf{Gain vs. best prior} & \best{+2.58~dB} & \best{+8.3\%} & \best{+37.7\%} & \best{+33.0\%} \\
\bottomrule
\end{tabular}
\end{table}

\begin{figure}[!t]
\centering
\includegraphics[width=1\linewidth]{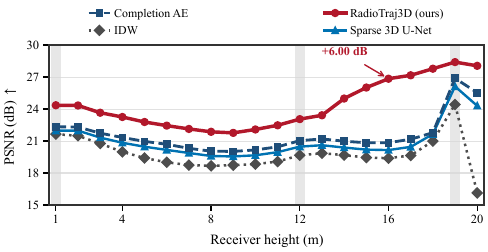}
\caption{Height-wise PSNR with observed-layer shading.}
\label{fig:heightwise_method_comparison}
\end{figure}

Under the matched transmitter-agnostic protocol, RadioTraj3D attains the best value for every metric in Table~\ref{tab:external_baselines}. Against Completion AE, the strongest matched baseline, it gains 2.58~dB in PSNR and 8.3\% in SSIM while reducing NMSE by 37.7\% and RMSE by 33.0\%. Figure~\ref{fig:qualitative_baselines} shows that these aggregate gains correspond to sharper shadow boundaries and more faithful broad propagation patterns. The height-wise diagnostic in Fig.~\ref{fig:heightwise_method_comparison} further shows a gain at every layer, including 6.00~dB over Completion AE at the unobserved layers from $z=16$ to 18~m.

To determine whether the gain depends on the outdoor training grid, we evaluate 3DiRM3200~\cite{rao2026r2net}, which contains 16 heights and 200 indoor layouts. No dataset-specific transmitter data are introduced, and only geometric anchors and distances are rescaled to its 8-cm spacing.

\begin{table}[!t]
\centering
\caption{Cross-dataset transfer on 3DiRM3200.}
\label{tab:external_3dirm3200}
\ninept
\setlength{\tabcolsep}{3pt}
\renewcommand{\arraystretch}{0.9}
\begin{tabular}{@{}lccc@{}}
\toprule
Method / height rule & PSNR $\uparrow$ & SSIM $\uparrow$ & NMSE $\downarrow$ \\
\midrule
Completion AE + Fixed & 31.90 & 0.9242 & 0.01252 \\
Completion AE + Adaptive & 31.99 & 0.9283 & 0.01211 \\
IDW + Adaptive & 30.80 & \second{0.9364} & 0.01603 \\
\rowcolor{oursbg}\textbf{RadioTraj3D + Adaptive} & \best{34.77} & \best{0.9578} & \best{0.00704} \\
\midrule
\rowcolor{gray!12}\textbf{Gain vs. best prior} & \best{+2.78~dB} & \best{+2.3\%} & \best{+41.9\%} \\
\bottomrule
\end{tabular}
\end{table}

After this geometric rescaling, Table~\ref{tab:external_3dirm3200} compares the fixed middle-layer rule at $z=8$ with the geometry-adaptive rule at $z_\Delta=3$. Adaptive height improves SSIM by 0.0041 and reduces NMSE by 0.000412, with layout-clustered 95\% intervals excluding zero. RadioTraj3D then gains 2.78~dB in PSNR and reduces NMSE by 41.9\% over adaptive Completion AE. These gains carry across environment type and vertical grid through both height selection and dual-view reconstruction.

\begin{figure}[!t]
\centering
\setlength{\tabcolsep}{0.6pt}
\renewcommand{\arraystretch}{0.9}
\ninept
\begin{tabular}{@{}*{5}{c}@{}}
\shortstack{Ground\\truth} & Ours & \shortstack{Sparse\\3D U-Net} & \shortstack{Completion\\AE} &
\shortstack{DAT-\\UNet} \\
\multicolumn{5}{c}{\ninept $z=1$ m} \\
\includegraphics[width=0.18\linewidth]{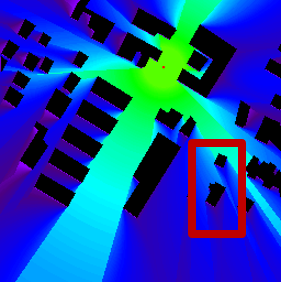} &
\includegraphics[width=0.18\linewidth]{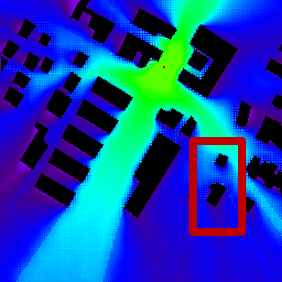} &
\includegraphics[width=0.18\linewidth]{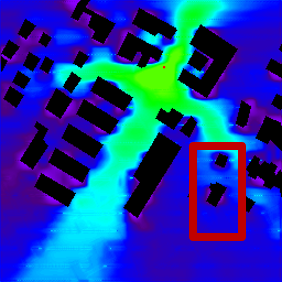} &
\includegraphics[width=0.18\linewidth]{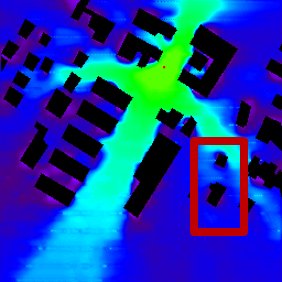} &
\includegraphics[width=0.18\linewidth]{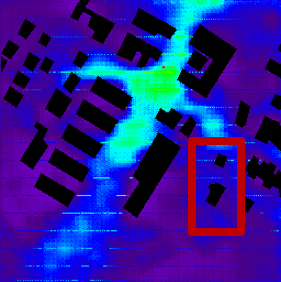} \\
\multicolumn{5}{c}{\ninept $z=12$ m} \\
\includegraphics[width=0.18\linewidth]{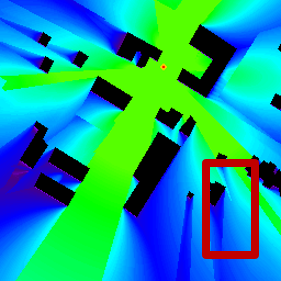} &
\includegraphics[width=0.18\linewidth]{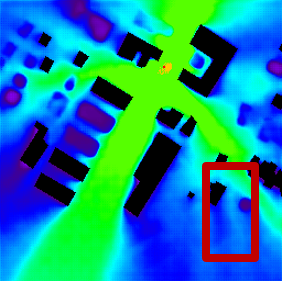} &
\includegraphics[width=0.18\linewidth]{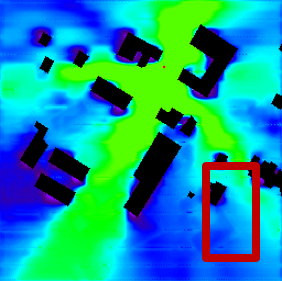} &
\includegraphics[width=0.18\linewidth]{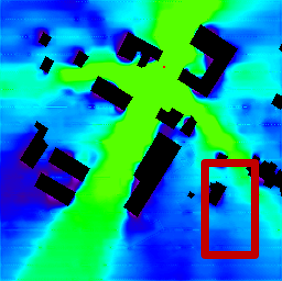} &
\includegraphics[width=0.18\linewidth]{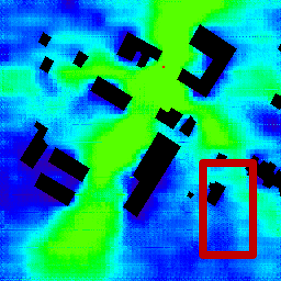} \\
\multicolumn{5}{c}{\ninept $z=20$ m} \\
\includegraphics[width=0.18\linewidth]{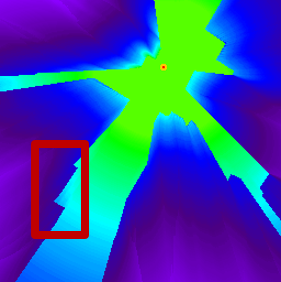} &
\includegraphics[width=0.18\linewidth]{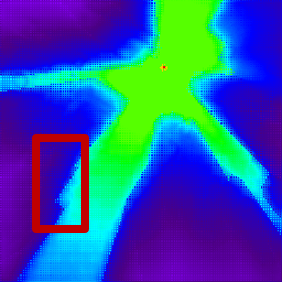} &
\includegraphics[width=0.18\linewidth]{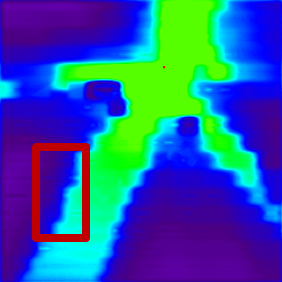} &
\includegraphics[width=0.18\linewidth]{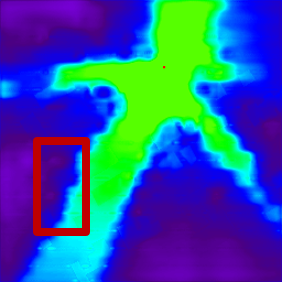} &
\includegraphics[width=0.18\linewidth]{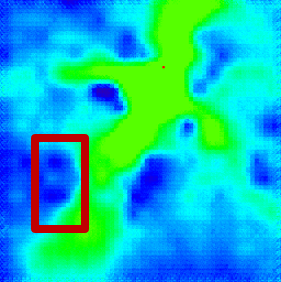}
\end{tabular}
\caption{Matched qualitative comparison on BID 609.}
\label{fig:qualitative_baselines}
\end{figure}

\subsection{RadioTraj3D Component Analysis}

\begin{table}[!h]
\centering
\caption{Fusion-rule ablation on UrbanRadio3D.}
\label{tab:radiorecon_components}
\ninept

\setlength{\tabcolsep}{2.1pt}
\renewcommand{\arraystretch}{0.9}
\begin{tabular}{@{}lccccc@{}}
\toprule
Fusion rule & Params & PSNR $\uparrow$ & SSIM $\uparrow$ & NMSE $\downarrow$ & RMSE $\downarrow$ \\
\midrule
3D branch only & 0 & 23.15 & 0.779 & 0.0289 & 0.0716 \\
Full-height only & 0 & 23.35 & 0.786 & 0.0278 & 0.0702 \\
Equal, $A=0.5$ & 0 & 23.55 & 0.793 & 0.0265 & 0.0688 \\
Learned global & 1 & 23.58 & 0.794 & 0.0263 & 0.0686 \\
Fixed $A=0.6$ & 0 & \second{23.60} & \second{0.794} & \second{0.0262} & \second{0.0684} \\
\rowcolor{oursbg}\textbf{Height-wise, ours} & \textbf{20} & \best{23.86} & \best{0.805} & \best{0.0247} & \best{0.0664} \\
\bottomrule
\end{tabular}
\end{table}

Figure~\ref{fig:heightwise_method_comparison} shows that relative branch accuracy changes with height, and Table~\ref{tab:radiorecon_components} tests how fusion should respond. Equal fusion improves on either branch, whereas one global weight brings little additional gain. From the displayed means, height-wise fusion improves learned global fusion by 0.28~dB in PSNR and 0.011 in SSIM. The paired differences for all four metrics have 95\% intervals excluding zero, and 38--41 of the 50 scenarios improve. The 20 coefficients add 4.20~ms, or 1.8\%, latency on an RTX~4090.

\subsection{Trajectory-Constrained Sampling Results}

\begin{table}[!h]
\centering
\caption{Sampler cross-evaluation at $N=500$ per height.}
\label{tab:sampling_analysis}
\ninept
\setlength{\tabcolsep}{3.5pt}
\renewcommand{\arraystretch}{0.9}
\begin{tabular}{@{}lccc@{}}
\toprule
Train & \multicolumn{3}{c}{Test sampler, PSNR / SSIM} \\
\cmidrule(lr){2-4}
 & Safe & Coverage & G-METS \\
\midrule
Safe & 20.72 / 0.7300 & \second{20.83 / 0.7330} & \best{21.04 / 0.7400} \\
Coverage & 20.82 / 0.7340 & \second{20.92 / 0.7360} & \best{21.14 / 0.7430} \\
G-METS & 20.84 / 0.7350 & \second{20.95 / 0.7370} & \best{21.22 / 0.7460} \\
\bottomrule
\end{tabular}
\end{table}

\begin{table}[!h]
\centering
\caption{G-METS design ablation at $N=500$ per height.}
\label{tab:gmets_design}
\ninept
\setlength{\tabcolsep}{3.2pt}
\renewcommand{\arraystretch}{0.9}
\begin{tabular}{@{}llcc@{}}
\toprule
Factor & Variant & PSNR $\uparrow$ & SSIM $\uparrow$ \\
\midrule
\rowcolor{oursbg}Sampler & \textbf{Full G-METS} & \best{21.22} & \best{0.7460} \\
Sampler & without spacing & 20.42 & 0.7165 \\
Sampler & without edge quota & 20.88 & 0.7366 \\
Sampler & without corridor quota & \second{20.92} & \second{0.7382} \\
\midrule
\rowcolor{oursbg}Height & \textbf{Adaptive G-METS} & \best{21.22} & \best{0.7460} \\
Height & Fixed 1, 12, 19~m & 20.87 & 0.7329 \\
Height & Fixed 1, 13, 19~m & \second{20.89} & 0.7334 \\
Height & Fixed 1, 14, 19~m & 20.77 & \second{0.7336} \\
\bottomrule
\end{tabular}
\end{table}

\begin{figure}[!h]
\centering
\includegraphics[width=0.7\linewidth]{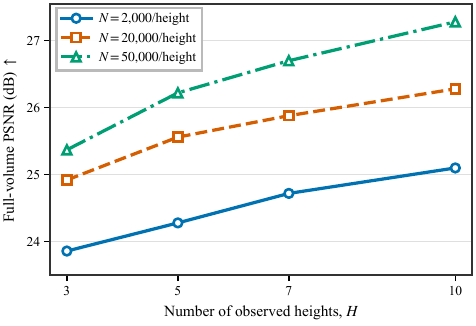}
\vspace{-1mm}
\caption{PSNR versus observed heights and sampling density.}
\vspace{-1mm}
\label{fig:height_budget_psnr}
\end{figure}

\begin{figure}[!t]
\centering
\setlength{\tabcolsep}{0.25pt}
\renewcommand{\arraystretch}{0.95}
\begin{tabular}{@{}c*{5}{c}@{}}
& \ninept $z=1$
& \ninept $z=4$
& \ninept $z=8$
& \ninept $z=10$
& \ninept $z=20$ \\

\ninept Truth
& \includegraphics[width=0.175\linewidth]
  {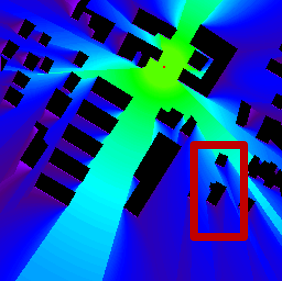}
& \includegraphics[width=0.175\linewidth]
  {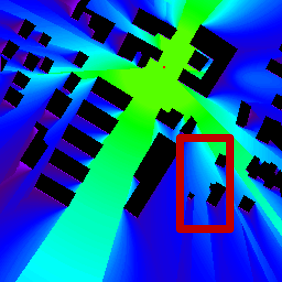}
& \includegraphics[width=0.175\linewidth]
  {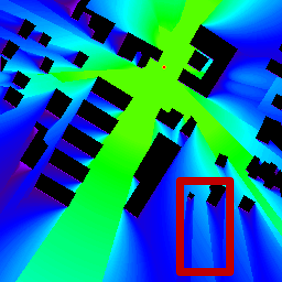}
& \includegraphics[width=0.175\linewidth]
  {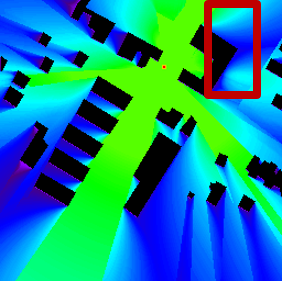}
& \includegraphics[width=0.175\linewidth]
  {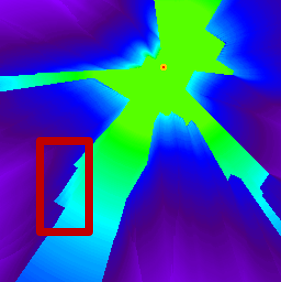} \\

\ninept Ours
& \includegraphics[width=0.175\linewidth]
  {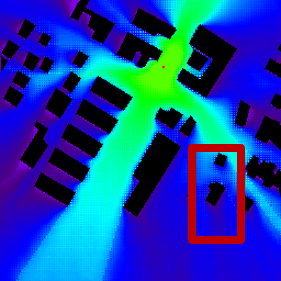}
& \includegraphics[width=0.175\linewidth]
  {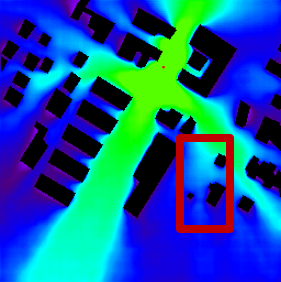}
& \includegraphics[width=0.175\linewidth]
  {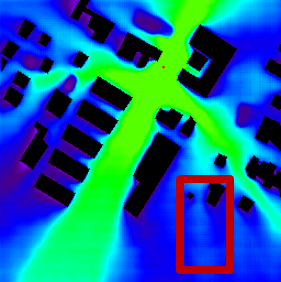}
& \includegraphics[width=0.175\linewidth]
  {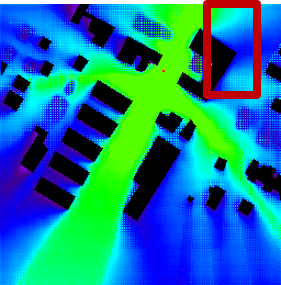}
& \includegraphics[width=0.175\linewidth]
  {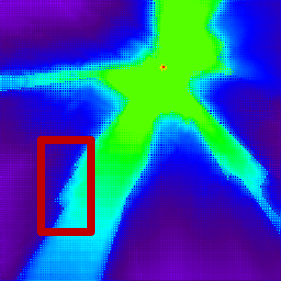}
\end{tabular}

\caption{Dense three-height reconstruction on BID 609.}
\label{fig:n20000_reconstruction}
\end{figure}

Across the three frozen training samplers, Table~\ref{tab:sampling_analysis} shows that G-METS test measurements yield the highest PSNR and SSIM in every row. Relative to the individual baselines, the PSNR gain ranges from 0.21 to 0.38~dB. Table~\ref{tab:gmets_design} attributes this gain to spacing control, regional allocation, and adaptive height selection. Sampler ablations cost up to 0.80~dB, and adaptive heights improve PSNR by at least 0.33~dB and SSIM by at least 0.0124 over every fixed-height alternative. Distributing 20,000 measurements across 10 heights uses one third as many measurements as the 60,000-point three-height setting while increasing PSNR from 24.92 to 25.10~dB in Fig.~\ref{fig:height_budget_psnr}. In a separate dense-plane diagnostic, complete planes at $z=1,12,19$~m produce 29.30~dB over the full volume and 28.92~dB on 17 unobserved heights. Figure~\ref{fig:n20000_reconstruction} confirms vertical reconstruction rather than plane copying.

\subsection{Real-World Validation on LACN AAV Measurements}

The final evaluation tests the reconstructor on field measurements. LACN-Before~\cite{huang2026lacn} contains AAV-collected reference signal received power (RSRP) measurements at 4.9~GHz and heights of 75, 230, 400, and 500~m. For each of three disjoint-region rotations, all models initialize from UrbanRadio3D, adapt for 800 updates on two regions, and predict a complete height withheld from the third. Table~\ref{tab:lacn_measured} aggregates 9,290 measured points under this protocol. Because LACN provides no registered building mask, every method receives a zero building channel. This test jointly changes frequency, height scale, data source, and geometry input.

\begin{table}[!h]
\centering
\caption{Performance evaluation on real-world measured dataset.}
\label{tab:lacn_measured}
\ninept
\setlength{\tabcolsep}{5.2pt}
\renewcommand{\arraystretch}{0.91}
\begin{tabular}{@{}lcc@{}}
\toprule
Method & PSNR $\uparrow$ & NMSE $\downarrow$ \\
\midrule
Completion AE~\cite{teganya2022deepcompletion} & \second{25.19} & \second{0.00671} \\
RadioDUN~\cite{chen2026radiodun} & 14.37 & 0.08097 \\
DAT-UNet~\cite{liu2026datunet} & 23.53 & 0.00984 \\
\rowcolor{oursbg}\textbf{RadioTraj3D} & \best{26.03} & \best{0.00553} \\
\midrule
\rowcolor{gray!12}\textbf{Gain vs. best prior} & \best{+0.84~dB} & \best{+17.5\%} \\
\bottomrule
\end{tabular}
\end{table}

The RadioTraj3D reconstructor achieves 26.03~dB in PSNR and 0.00553 in NMSE. Relative to Completion AE, the strongest baseline, it gains 0.84~dB and reduces NMSE by 17.5\%. The PSNR gain is positive at every withheld height and reaches 2.44~dB at 400~m, establishing a consistent advantage across the measured height range. The same ordering holds in physical units, where RadioTraj3D reduces mean absolute error (MAE) from 3.12 to 2.76~dB and RMSE from 4.13 to 3.75~dB. These measured-data results confirm that the dual-view reconstructor transfers vertical information even without a registered building mask.

\FloatBarrier

\section{Conclusion}
\label{sec:conclusion}

RadioTraj3D coordinates feasible sampling and full-volume reconstruction through height. G-METS targets informative transitions under collision and budget constraints, and dual-view calibration balances local 3D structure with full-height context using $D$ parameters. Matched comparisons, 3DiRM3200 transfer, and real-world LACN adaptation demonstrate component effectiveness and transmitter-agnostic generalization beyond outdoor training.

\FloatBarrier
\clearpage
\section*{Acknowledgment}
The authors used generative AI to translate the Chinese draft of this paper into English.

\bibliographystyle{IEEEbib}
\bibliography{refs}

\end{document}